# Probing macroscopic temperature changes with non-radiative processes in hyperbolic meta-antennas

Nils Henriksson[1], Alessio Gabbani[2,3], Gaia Petrucci[2], Denis Garoli[4,5], Francesco Pineider[2,3], and Nicolò Maccaferri[1,6]*

[1]Department of Physics, Umeå University, Linnaeus väg 24, 901 87 Umeå, Sweden

[2] Department of Chemistry and Industrial Chemistry, University of Pisa, via Moruzzi 13, 56124, Pisa (Italy)

[3] Department of Physics and Astronomy, University of Florence, via Sansone 1, 50019, Sesto Fiorentino (Italy)

[4] Istituto Italiano di Tecnologia, Via Morego 30, 16163 Genova (Italy)

[5] Dipartimento di Scienze e Metodi dell'Ingegneria, University of Modena and Reggio-Emilia, Via Amendola 2, 42122 Reggio Emilia (Italy)

[6] Umeå Centre for Microbial Research, 901 87 Umeå, Sweden

*nicolo.maccaferri@umu.se

## Abstract

Multilayered metal-dielectric nanostructures display both strong plasmonic behavior and hyperbolic optical dispersion. The latter is responsible for the appearance of two separated radiative and non-radiative channels in the extinction spectrum of these structures. This unique property can open a wealth of opportunities towards the development of multifunctional systems that simultaneously can behave as optimal scatterers and absorbers at different wavelengths, an important feature to achieve multiscale control light-matter interactions in different spectral regions for different types of applications, such as optical computing or detection of thermal radiation. Nevertheless, the temperature dependence of the optical properties of these multilayered systems has never been investigated. In this work we study how radiative and non-radiative processes in hyperbolic meta-antennas can probe temperature changes of the surrounding medium. We show that, while radiative processes are essentially not affected by a change in the external temperature, the non-radiative ones are strongly affected by a temperature variation. By combining experiments and temperature dependent effective medium theory, we find that this behavior is connected to enhanced damping effects due to electron-phonon scattering. Contrary to standard plasmonic systems, a red-shift of the non-radiative mode occurs for small variations of the environment temperature. Our study shows that to probe temperature changes it is essential to exploit non-radiative processes in systems supporting plasmonic excitations, which can be used as very sensitive thermometers via linear absorption spectroscopy.

## Introduction

Multilayered metal-insulator structures represent an interesting platform for engineering both the spatial and temporal properties of the electric permittivity in photonic devices[1]. These structures offer the possibility to bring the refractive index close to zero[2–4], enabling novel optical phenomena such as perfect transmission through distorted waveguides[5], cloaking[6,7] and inhibited diffraction[8]. They also present an almost infinite density of states[9] and are widely used for nanoscale light confinement and guiding[10–14], as well as manipulating scattering, absorption and nonreciprocal propagation of light[15–22], generating optical vortex beams[23] and tailoring optical nonlinearities[24–30] and optical magnetism[31,32], and for highly-sensitive detection[33–40] and ultrafast all-optical switching[41–44]. In addition, metal-insulator multilayers display hyperbolic optical dispersion[45–53], and thanks to this property they have successfully been implemented as negative index materials[54–56], color filters[57], quantum yield enhancers[58–62], super-absorbers driving resonant gain singularities[63–66], as well as for hot-electron generation and manipulation[67,68], super resolution imaging[69], ultra-compact optical quantum circuits[70] and lasing[71].

In this context, it has been shown that multilayered metal-dielectric antennas displaying hyperbolic dispersion have two separated radiative and non-radiative channels[19], and that this property can be exploited to manipulate electron dynamics on ultrafast timescales[68] as well as for practical applications such as localized hyperthermia[72] and enhanced spectroscopy[21]. The unique property of having two separated spectral regions where either a radiative or a non-radiative process is dominating on the other, and vice versa, can open a plenty of opportunities in developing multifunctional systems which can behave at the same time as optimal scatterers and absorbers. This is an essential property to achieve a multiscale control light-matter interactions in different spectral regions for different types of applications, such as optical computing or detection of thermal radiation. Nevertheless, the temperature dependence of the optical properties of these multilayered systems, has never been investigated. In this work we study how radiative and non-radiative channels in hyperbolic meta-antennas can probe temperature changes of the embedding medium. We experimentally show that, while radiative modes are essentially not affected by a change in the external temperature, the non-radiative channel is strongly affected by a temperature variation, displaying a reduction of the absorption cross section together with a broadening/red-shift of the resonance bandwidth/peak. By combining effective medium theory and a temperature-dependent Drude-Lorentz model, we show that this behavior is mainly connected to the damping inside the meta-antennas due to electron-phonon scattering followed by a red-shift of the plasma frequency of the metallic contribution to the permittivity of the metal-dielectric multilayered systems. Our findings also show that, contrary to standard plasmonic systems, a red-shift of the non-radiative mode occurs for relatively small (of the order of few degrees) variations of the environment temperature. Thus, our study sheds new light on the physics of how radiative and non-radiative channels can probe temperature changes and shows that hyperbolic structures can be used as very sensitive thermometers if we track the spectral variation of their non-radiative processes via linear absorption spectroscopy.

## Results and discussion

A surface plasmon polariton (SPP) corresponds to a light-driven collective oscillation of electrons localized at the interface between materials with dielectric ($\varepsilon > 0$) and metallic ($\varepsilon < 0$) dispersions. If the interface is flat, as in a thin layer, propagating SPP can propagate along the interface. When multiple metal/dielectric interfaces supporting SPPs occur within subwavelength separation, the associated coupled electromagnetic field exhibits a collective behavior, which can be modeled by an effective medium approximation (EMA) and the dispersion relation presents a unique anisotropic optical dispersion. More precisely, an effective permittivity tensor $\hat{\varepsilon}$ can be derived such as

$$\hat{\varepsilon} = \begin{pmatrix} \varepsilon_\parallel & 0 & 0 \\ 0 & \varepsilon_\parallel & 0 \\ 0 & 0 & \varepsilon_\perp \end{pmatrix}$$

with $\varepsilon_\perp$ ($\varepsilon_\parallel$) the perpendicular (parallel) component with respect to the meta-antennas plane, satisfying $\varepsilon_\perp \varepsilon_\parallel < 0$ and thus presenting a iso-frequency surface with a hyperbolic shape[45].

In our study, we investigate the linear optical response of hyperbolic meta-material (HMM) antennas, known also as hyperbolic meta-antennas. The optical extinction, which accounts for of both scattering (radiative) and absorption (non-radiative) processes, can be described by the EMA (Figure 1a), at different temperatures. The structures are disk-shaped cylinders made of five layers of Au and $TiO_2$ with thickness of 10nm and 20nm, respectively, and a nominal radius of 125 nm. When illuminated by light, localized plasmon resonances (LSPRs), collective oscillations of the free electron cloud driven by the electromagnetic field of the incident light, occur in specific spectral regions (Figure 1a). The extinction of the structures is then greatly enhanced at the wavelengths corresponding to the LSPR excitations. It has been shown that hyperbolic meta-antennas exhibit a more complex plasmonic response compared to Au antennas with the same geometrical shape and size, featuring two spectrally separated scattering and absorbing modes[19] (see top panel in Figure 1a). However, when the ambient temperature surrounding the structure increases, the refractive index of the meta-antenna changes, and thus also the way these structures can interact with light, in particular how they absorb radiation (Figure 1b). The main reason for this change is connected to the fact that the metal building block of our meta-antennas is sensitive to temperature changes, and this sensitivity is amplified by the fact that light-matter interactions at the non-radiative mode are greatly increased[19,68,72]. Moreover, the size of the antennas increases due to thermal expansion, which in turn lowers the electron density, and thus the plasma frequency[73], causing a change in the Au permittivity. While a decrease of the effective mass has been shown to counteract the effect of the lower electron density on the plasma frequency of thin Au films[73] causing the plasma frequency to increase, we see that our system experiences a clear decrease of the plasma frequency. To investigate the temperature dependence of the optical response of the hyperbolic meta-antennas, we performed linear absorption spectroscopy measurements to characterize the extinction as a function of wavelength at different ambient temperatures (see Methods and Supplementary Information for details about the experiments and the experimental setup, respectively).

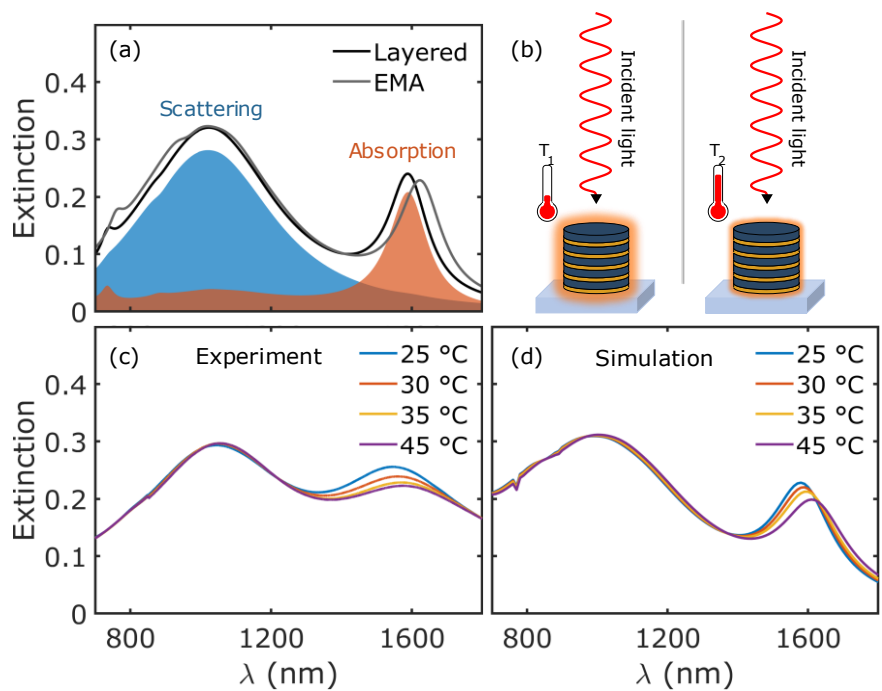

**Figure 1.** (a). Calculated extinction cross section for a hyperbolic meta-antenna of diameter 250 nm and 5x Au(10 nm)/TiO$_2$(20 nm) layers thickness. The blue and orange curves are the absorption and scattering contributions to the total extinction (black line). The extinction of the structure calculated using the EMA is also plotted (grey curve). (b) Schematic of the experiment showing the decreased absorption (glowing red area surrounding the structure) at increased surrounding temperature. Experimental (c) as well as simulated (d) extinction at different temperatures. Extinction spectra at four different temperatures have been selected to highlight the main change in extinction as the surrounding temperature increases.

As illustrated in Figure 1c, the extinction of our meta-antennas decreases at the absorption mode (thus where non-radiative processes dominate) with increasing temperature. At the same time, the extinction at the scattering mode remains almost unchanged for the temperature variation range we are considering here, that is from room temperature (25 °C) up to 45 °C. In addition to a decrease of extinction at the absorption mode, we can also appreciate a broadening and a red-shift of the resonance. The first effect can be explained by considering an increased damping of the electrons inside the metal, while the red-shift needs a more detailed analysis.

To understand the physics underlying the temperature-dependence behavior of both non-radiative and radiative processes in hyperbolic meta-antennas, we performed finite element method (FEM) simulations using the commercial software COMSOL Multiphysics, utilizing the wave optics module to calculate the steady state extinction as a function of the incident light wavelength (see Figure 1d). The system was modeled using the EMA, with the permittivity components calculated as[22,74]

$$\varepsilon_\parallel = \frac{t_d \varepsilon_m + t_m \varepsilon_d}{t_d + t_m} \tag{1}$$

$$\varepsilon_\perp = \frac{\varepsilon_m \varepsilon_d (t_d + t_m)}{t_d \varepsilon_m + t_m \varepsilon_d} \tag{2}$$

where $\varepsilon_{m,(d)}$ is the Au (TiO$_2$) permittivity, and $t_{m,(d)}$ is the layer thickness of Au (TiO$_2$). In our study, we focus on wavelengths where intraband transitions are the main contribution to the permittivity[22], thus we used a Drude model to describe Au permittivity[75]. In Figure 2a, we plot the in-plane component of the hyperbolic permittivity, $\varepsilon_\parallel$, at two temperatures, 25 °C and 45 °C. As it can be inferred by looking at the figure, the real part (blue curves) is negative, thus indicating that in the plane the hyperbolic meta-antennas have a metallic behavior. We can see that this part is strongly affected by a change in the surrounding medium temperature, while the imaginary part (orange curves) is less affected by temperature change. It is worth noticing that the refractive index of TiO$_2$ is not constant in the wavelengths range considered (see dashed blue curve in Figure 2b)[76]. Nevertheless, for the dielectric material in our EMA model, we used a constant refractive index of 2 to match the experimental extinction peaks. In Figure 2b we plot both the experimental refractive index of TiO$_2$ from Ref. [[76]] and the one calculated using our EMA and using the out-of-plane component $\bar{\varepsilon}_\perp$ assuming a constant value of 2 for the TiO$_2$ refractive index. The two values are very similar and the imaginary part of the refractive index, the loss coefficient, is almost zero over the whole spectral range. This indicates that our out-of-plane refractive index is the refractive of an insulator, and the temperature variation of $TiO_2$ refractive index with temperature is known to be very negligible in the range of temperatures we are considering here[77–80].

Thus, we can disregard contributions to the optical extinction due to the temperature dependence of the out-of-plane component of our hyperbolic permittivity and consider only the contributions from the in-plane and metallic component $\tilde{\varepsilon}_{\parallel}$. In other words, we considered a Drude model to describe $\varepsilon_{\parallel}$, the metallic-like permittivity component, whilst using the dielectric-like component $\varepsilon_{\perp}$ as described in Eq. 2, such that

$$\tilde{\varepsilon}_{\parallel} = 1 - \frac{\omega_p^2}{\omega(\omega + i\Gamma)}, \tag{3}$$

where $\omega_p$ is the plasma frequency and $\Gamma$ the damping constant at room temperature. We made a fit of Eq. 3 to the in-plane component $\varepsilon_{\parallel}$ in Eq. 1, which is based on a Drude model of the Au permittivity[75]. The values of the parameters were found to be $\omega_p = 4.937$ eV and $\Gamma = 0.0184$ eV. However, since interband transitions are not considered in this model, the value of the damping is understated. To account for this, we increased the damping to $\Gamma = 0.15$ eV, which well reproduced the spectra from the initial EMA model in Figure 1a, which in turn is based on the Brendel-Bormann permittivity model on Au.[81] This approach was used to enable precise modelling of the temperature effects on both plasma frequency and damping constant. We thus built a temperature dependent model for $\tilde{\varepsilon}_{\parallel}$ following previous works done on metals[73,82]. Following similar procedures, in our model we derived the temperature dependences of plasma frequency and damping constant as

$$\widetilde{\omega}_p(T) = \frac{\omega_p}{\sqrt{1 + \gamma(T - T_0)}} \tag{4}$$

$$\tilde{\Gamma}(T) = \Gamma + \Gamma_{e-ph}(T) - \Gamma_{e-ph}(T_0), \tag{5}$$

where $T_0$ is the room temperature and $\omega_p$ the plasma frequency, $\gamma$ the thermal expansion coefficient and $\Gamma$ the damping used in Eq. 3. The temperature-dependent damping due to electron-phonon scattering at temperatures much larger than the Debye temperature of Au, $\theta_D = 170$ K, can be calculated as[73]

$$\Gamma_{e-ph}(T) = \Gamma_0\left(\frac{2}{5} + \frac{T}{\theta_D}\right)$$

The thermal expansion coefficient $\gamma$ and the electron-phonon scattering rate $\Gamma_0$ were used as fitting parameters to match the behavior of the experimental results. We found that $\gamma = 4 \cdot 10^{-3}$ K$^{-1}$ and $\Gamma_0 = 0.1$ eV gave the best fit. Inserting Eqs. 4 and 5 into Eq. 3 provided us with a temperature dependent metallic-like permittivity $\tilde{\varepsilon}_{\parallel}$. The temperature-dependent real and imaginary parts of $\tilde{\varepsilon}_{\parallel}(\omega, T)$ are displayed in Figure 2a. We only considered thermal effects on the plasma frequency and the electron-phonon induced damping as they are the main contributors to the change in permittivity, as previously shown for pure metallic structures[82], where both electron-electron and phonon-phonon scattering processes can be ignored. The temperature dependence of both quantities is displayed in Figure 3a. By comparing the experimental results (Figure 1c) with our simulations (Figure 1d), it is clear that our simple model replicates the behavior of the extinction spectrum when the ambient temperature increases, thus a more complex model that includes the electron-electron, phonon-phonon and even electron-surface scattering-induced damping is not necessary (more details about the simulations of absorption and scattering cross sections can be found in Methods).

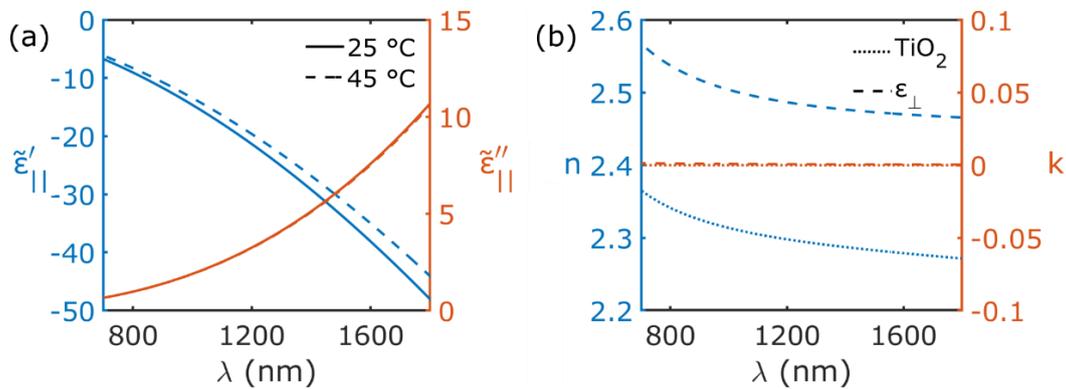

**Figure 2.** (a) The temperature-dependent complex permittivity $\tilde{\varepsilon}_{\parallel}$ at 25 °C and 45 °C. (b) The experimental refractive index of TiO$_2$[76] (dotted curve) and the refractive index calculated using $\tilde{\varepsilon}_{\perp}$ (dashed curve) assuming a constant value of 2 for the TiO$_2$ refractive index.

A full evaluation and comparison between the experimental and the numerical results at the absorption mode were performed by computing the relative temperature-induced change of the full width at half maximum (FWHM) (Figure 3b), the spectral shift of the resonance peak (Figure 3c) as well as the change of peak extinction (Figure 3d). This was done using a double Lorentzian fit (see Supplementary Information for more details). The figure shows that our model can predict the change in FWHM accurately, while the red-shift as well as the decrease in magnitude show similar changes between 25°C and 45°C, albeit with a more linear behavior than the experimental counterpart.

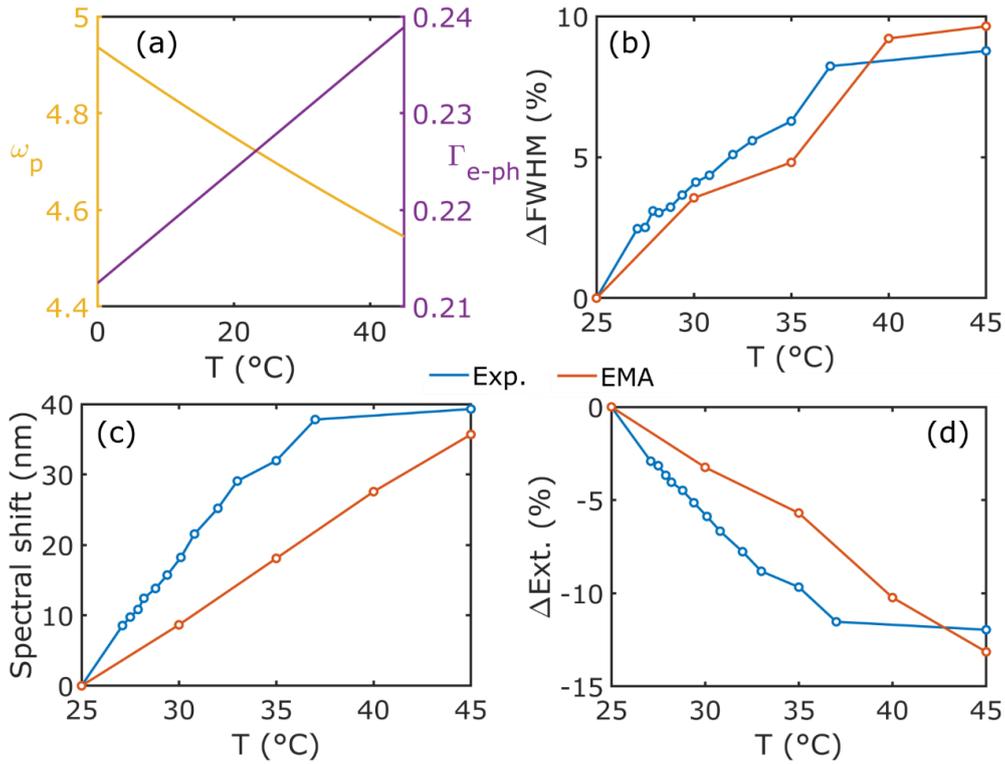

**Figure 3.** (a) The temperature dependent plasma frequency and damping in eV. Effects of changing surrounding medium temperature on (b) FWHM, (c) spectral position of extinction peak, and (d) peak extinction intensity, for both the experimental (blue curves) and the simulated (orange curves) cases.

It is worth mentioning here that, given that the refractive index of $TiO_2$[76] remains almost constant and that $Au$[81] permittivity can be described by the Drude model between $\lambda = 1$ μm and $\lambda = 5$ μm, this implies that out model is valid further into the infrared spectrum. Thus, we simulated hyperbolic meta-antennas where we increased the radius of the structures to red-shift both peaks of the extinction spectrum and investigate whether our system behaves in the same way also when both radiative and non-radiative processes move towards the mid infrared. In Figure 4, we show the extinction change at the scattering and absorption modes between 25 °C and 45 °C for structures with radii 130 nm, 180 nm, and 230 nm. While the absorption mode is still very sensitive to temperature changes, the scattering mode is still not affected so much by a temperature variation around room temperature. Thus, our hyperbolic meta-antennas can be used as both sensitive thermometers of macroscopic temperature changes if non-radiative processes are optically detected, for instance via linear absorption spectroscopy, and can be used for thermal sensing in the infrared spectral region. Finally, we want to highlight that it is not possible to achieve this feature with standard plasmonic antennas made of gold and with the same geometrical parameters such as they resonate in these spectral range (see Figure 4). This is due to the fact that scattering is the dominant extinction pathway in these structures.

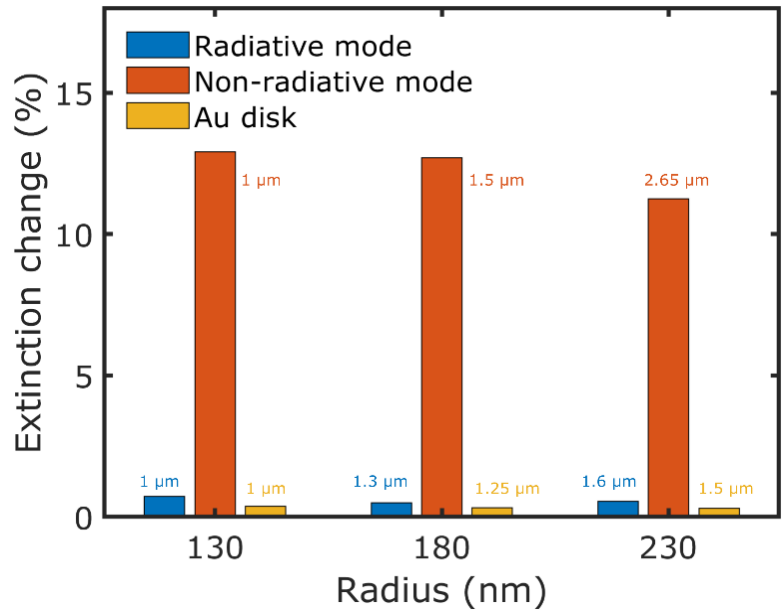

**Figure 4.** Relative extinction changes between 25 °C and 45 °C at the scattering and absorbing mode for different-sized antennas, calculated with the EMA model, as well as the corresponding extinction change of a gold disk of the same radius as the HMM structure with a height of 50 nm. The simulation of the Au disk was performed using the Brendal-Bormann permittivity for Au[81]. The wavelength of each resonant mode is displayed in the figure.

In summary, we have experimentally studied how hyperbolic meta-antennas linear optical response changes by varying the surrounding medium temperature from room temperature up to 45 °C. We found that the non-radiative contribution (absorption) to the extinction is sensitive to temperature changes, while the radiative contribution (scattering) is not sensitive in the range of temperatures considered. By combining effective medium approximation, and temperature-dependent Drude model, we showed that the main effects affecting the linear optical response at the absorption peak are related to a drastic change of the electron damping and the plasma frequency due to electron-phonon scattering processes, and this effect results in a reduction of the absorption cross section together with a broadening/red-shift of the resonance width/peak. Thus, non-radiative processes in hyperbolic meta-antennas, compared to their counterpart in pure metallic plasmonic structures, are very sensitive to temperature changes and can be eventually used as temperature detectors in devices using optical read-out schemes. Thus, we envision that these types of nanostructured metamaterials can be implemented as thermal radiation sensors in future all-optical technologies.

## Methods

Fabrication. The multilayers pillars were fabricated following few steps of process: Electron Beam lithography was performed on MMA:PMMA(950K) bilayer spin coated at 3000rpm on the substrate; after the exposure and successive resist development in IPA:DI Water (2:1), electron-beam evaporation was used to deposit the different layers in the system. (The thickness of the different layers were first calibrated performing spectroscopic elipsometry on sample where a single layer was deposited). Finally, the lift-off was performed by immersing the sample in acetone for 30 minutes.

Experiments. UV-vis-NIR extinction spectra were acquired using a JASCO V670 commercial spectrophotometer working in transmission configuration. A metal ceramic heater element (HT19R, Thorlabs) was placed in contact with the sample substrate and used to heat the sample during the spectra. The ceramics has a 4 mm hole that allows light to pass through the sample. The sample temperature is controlled with a 100 Ohm resistive temperature sensor placed on the ceramics. The temperature was kept constant during each spectrum with maximum variations of 1 °C. Extinction spectra were acquired at temperatures between 25 and 45 °C.

Numerical calculations. The steady-state extinction spectra are calculated from FEM simulations in COMSOL Multiphysics on the full 150 nm high disk with a 130 nm radius, utilizing the EMT approximation. Using scattered field formulation in the frequency domain, the absorption and scattering cross-sections were calculated as

$$\sigma_{abs} = \frac{1}{I_0} \iiint Q \, \mathrm{d}V$$

$$\sigma_{sc} = \frac{1}{I_0} \iint (\boldsymbol{n} \cdot \boldsymbol{S}) \, \mathrm{d}s$$

where $I_0$ is the intensity of the incident light, $Q$ is the power absorbed by the structure, and $(\boldsymbol{n} \cdot \boldsymbol{S})$ is the Poynting vector in the normal direction of the surface of the particles. By dividing both quantities by the cross-sectional area of the computational domain with normal along the direction of the light, we obtained absorption, scattering, and therefore extinction values similar to the experimental results. An anisotropic permittivity tensor $\hat{\varepsilon}$ with a temperature dependent metallic-like component $\tilde{\varepsilon}_\perp(T)$ was used to account for the effect of the temperature change on the optical response of the meta-antennas.

## Acknowledgements

NH and NM acknowledge support from the Swedish Research Council (grant no. 2021-05784) and the European Innovation Council (grant n. 101046920). We acknowledge Attilio Zilli and Joel Kuttruff for fruitful discussions.

## Data availability statement

The authors declare that the data supporting the findings of this study are available within the paper, and its supplementary information files. The raw data are available from the corresponding author upon request.

**TOC Figure**

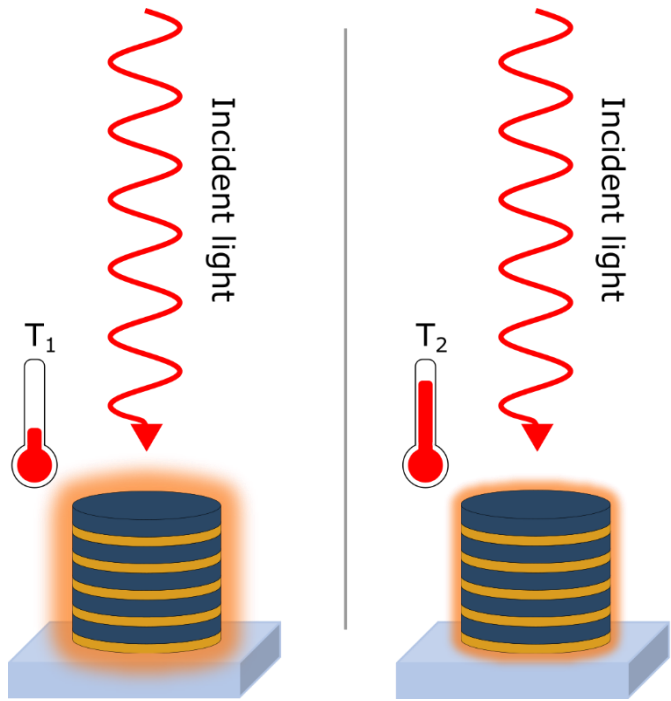
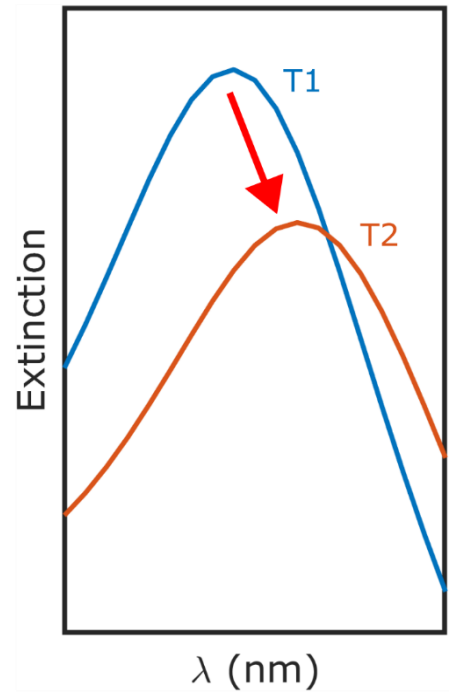

# Probing temperature changes using non-radiative and radiative processes in hyperbolic meta-antennas

**Nils Henriksson[1], Alessio Gabbani[2], Gaia Petrucci[2], Denis Garoli[3], Francesco Pineider[2], and Nicolò Maccaferri[1,4*]**

[1]Department of Physics, Umeå University, Linnaeus väg 24, 901 87 Umeå, Sweden

[2] University of Pisa

[3] Istituto Italiano di Tecnologia

[4] University of Modena and Reggio-Emilia

[5] Umeå Centre for Microbial Research, 901 87 Umeå, Sweden

*nicolo.maccaferri@umu.se

## 1. Calculating the FWHM

To compare the effect of the temperature in the experimental results (Figure 1c in the main paper) with the simulations in Figure 1d (main paper), we made a double Lorentzian fit on the different data

$$f(\lambda) = \frac{I_0}{\left[1 + \left(\frac{x - x_0}{\gamma_0}\right)^2\right]} + \frac{I_1}{\left[1 + \left(\frac{x - x_1}{\gamma_1}\right)^2\right]},$$

where $\{I_{0,1}, x_{0,1}, \gamma_{0,1}\}$ are fitting parameters. $I_{0,1}$ are the magnitudes of each peak, $x_{0,1}$ the position of the peaks and $2\gamma_{0,1}$ the full width at half maximum (FWHM) of each peak (see Figure S2).

## 2. EMA-approximation

In our work, we presented a new, simple method to numerically estimate the effect of a changing ambient temperature on multilayered nanoantennas. However, as a validation of the EMA approach for this system, we performed a simulation using the EMA with Au permittivity by Rakić et al.[1], and the anisotropic permittivity tensor implemented as described in the main text. We also made a simulation of a layered structure. The results are displayed in Figure 1a, showing a similar extinction spectrum for both cases.

**Supplementary figures**

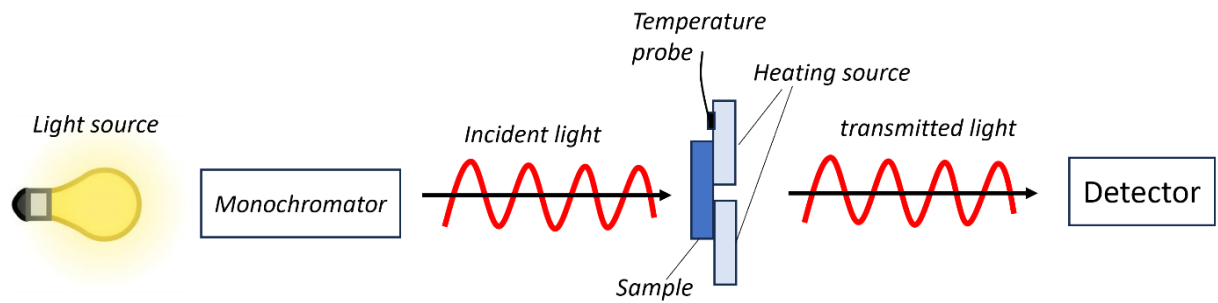

**Figure S1.** Experimental setup used in the experiments.

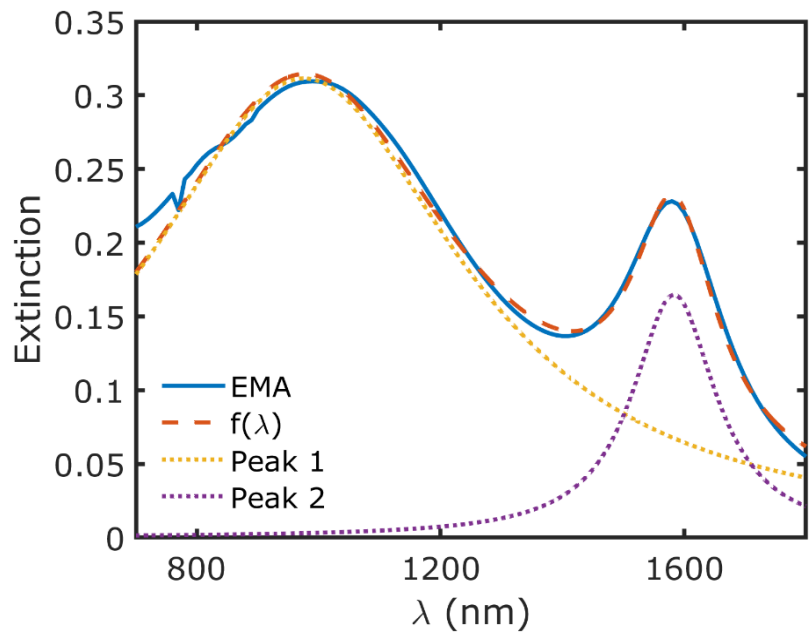

**Figure S2.** Example of a double Lorentzian fit. The dashed line shows the fit of the data, and the dotted curves show the fit of each peak.